\documentclass{aastex701}

\usepackage{textcomp}
\usepackage[normalem]{ulem}

\usepackage{amsmath}
\usepackage[utf8]{inputenc}
\usepackage{tabularx}

\newcommand\clearrow{\global\let\rowmac\relax}
\usepackage{booktabs}
\usepackage{multirow}
\usepackage{amsmath}
\usepackage{graphicx} 
\usepackage{hyperref}

\begin{document}

\title{Reconstruction of Solar EUV Irradiance Using CaII K Images 
and SOHO/SEM Data with Bayesian Deep Learning
and Uncertainty Quantification
}

\author[orcid=0000-0001-6460-408X,sname='Jiang']{Haodi Jiang} 
\affiliation{Institute for Space Weather Sciences, New Jersey Institute of Technology, University Heights, Newark, NJ 07102, USA}
\affiliation{Department of Computer Science, Sam Houston State University, Huntsville, TX 77341, USA}
\email[show]{haodi.jiang@shsu.edu}  

\author[orcid=0000-0002-3669-1830,sname='Li']{Qin Li}
\affiliation{Institute for Space Weather Sciences, New Jersey Institute of Technology, University Heights, Newark, NJ 07102, USA}
\affiliation{Center for Solar-Terrestrial Research, New Jersey Institute of Technology, University Heights, Newark, NJ 07102, USA}
\email{ql47@njit.edu}  

\author[orcid=0000-0002-2486-1097,sname='Wang']{Jason T. L. Wang}
\affiliation{Institute for Space Weather Sciences, New Jersey Institute of Technology, University Heights, Newark, NJ 07102, USA}
\affiliation{Department of Computer Science, New Jersey Institute of Technology, University Heights, Newark, NJ 07102, USA}
\email{wangj@njit.edu}

\author[orcid=0000-0002-5233-565X,sname='Wang']{Haimin Wang}
    \affiliation{Institute for Space Weather Sciences, New Jersey Institute of Technology, University Heights, Newark, NJ 07102, USA}
\affiliation{Center for Solar-Terrestrial Research, New Jersey Institute of Technology, University Heights, Newark, NJ 07102, USA}
\affiliation{Big Bear Solar Observatory, New Jersey Institute of Technology, 40386 North Shore Lane, Big Bear City, CA 92314, USA}
\email{wangha@njit.edu}

\author[orcid=0000-0002-4525-9038,sname='Criscuoli']{Serena Criscuoli}
\affiliation{National Solar Observatory, 3665 Discovery Drive, Boulder, CO 80303, USA}
\email{scriscuo@nso.edu} 

\correspondingauthor{Haodi Jiang}

\begin{abstract}
Solar extreme ultraviolet (EUV) irradiance plays a crucial role in heating the Earth's ionosphere, thermosphere, and mesosphere, affecting atmospheric dynamics over varying time scales. 
Although significant effort has been spent
studying short-term EUV variations from solar transient events,  
there is little work to explore
the long-term evolution of the EUV flux over multiple solar cycles.
Continuous EUV flux measurements have only been available since 1995, 
leaving significant gaps in earlier data.
In this study, we propose a Bayesian deep learning model, named SEMNet, to fill the gaps.
We validate our approach by applying SEMNet to
construct SOHO/SEM EUV flux measurements
in the period between 1998 and 2014
using CaII K images from the Precision Solar Photometric Telescope.
We then extend SEMNet through transfer learning
to reconstruct solar EUV irradiance
in the period between 1950 and 1960
using CaII K images from the
Kodaikanal Solar Observatory.
Experimental results show that
SEMNet provides reliable predictions along with uncertainty bounds, 
demonstrating the feasibility of CaII K images
as a robust proxy for long-term EUV fluxes. 
These findings contribute to a better understanding of solar influences on Earth's climate over extended periods.
\end{abstract}

\section{Introduction} 
\label{sec:intro}

Solar extreme ultraviolet (EUV) irradiance
is a major source of heating and
ionization in the upper atmosphere of the Earth
\citep{2011JGRA..116.1102L,Vourlidas2018}.
The solar EUV spectrum covers
the wavelength range from 1 to 120 nm, varying on timescales of
minutes, days, months, to multiple years in a solar cycle.
The variability of the EUV spectrum
is an important driver of space weather
\citep{2008AnGeo..26..269L,2014JSWSC...4A..30H}.
When solar activity is elevated, enhanced solar EUV driving causes
adverse space weather effects such as radio communication blackouts
\citep{doi:10.1126/sciadv.aaw6548, ISHII2024}.

The lack of continuous long-term observations in the EUV
is a long-standing problem in solar irradiance studies. 
Consistent EUV flux measurements have been achieved since 1995. The
longest coverage is offered by the Solar EUV Monitor (SEM) 
on board the Solar and Heliospheric Observatory (SOHO), 
which provides three EUV bands of measurements, 
covering from 1995 to the present
\citep{2002AdSpR..29.1963J}.
Later, the Solar EUV Experiment (SEE) 
on board the Thermosphere
Ionosphere Mesosphere Energetics and Dynamics (TIMED) satellite
measures solar spectral irradiance from 0.1 to 194 nm in 1-nm intervals
that are closely related to the energetics in the highly variable layers of the
Earth's atmosphere above 60 km
\citep{2005JGRA..110.1312W}.
More recently, the EUV Variability Experiment (EVE) on board the Solar Dynamics Observatory (SDO)
has measured solar EUV irradiance from 0.1 to 105 nm with unprecedented spectral resolution (0.1 nm) and
temporal cadence (ten seconds) since 2010
\citep{2012SoPh..275..115W}.
In addition, the Extreme Ultraviolet Sensor of the Geostationary Operational Environmental Satellite (GOES/EUVS) on board the GOES-R series satellites has measured solar irradiance on specific spectral lines from 5 to 127 nm since 2017 \citep{2009SPIE.7438E..04E}.

CaII K images provide proxies for the spatial
distribution of solar emission \citep{2008A&A...484..591C}. 
The relations between indices extracted from CaII K imagery or from disk-integrated CaII K emission, and solar irradiance, especially in the UV, have been pointed out in
several studies \citep[e.g.][]{1996GeoRL..23.2207W, 2000JApA...21..293K, 2010AN....331..696N, 2017ARA&A..55..159L, 2024JSWSC..14...34C}.
Such relations are used to develop models of solar irradiance \citep{2011JGRD..11620108F, 2018ApJ...865...22C, 2020SoPh..295...38B, 2024JSWSC..14....9C, 2024ApJ...976...11P},
thus allowing one to construct long and uniform time series of irradiance variability necessary for Earth climate studies \citep{2010RvGeo..48.4001G, gmd-10-4005-2017}.
Scaling relations between CaII K and UV/EUV emission are also found in other stars, and are of fundamental importance because UV/EUV measurements are quite rare for stars and strongly hampered by interstellar medium absorption \citep{2020A&A...644A..67S}.
Here, we propose using observed CaII K images to reconstruct solar EUV irradiance,
specifically SEM EUV fluxes. 
Our work is motivated by the research of \citet{Foukal1998}, 
who used digitized CaII K spectroheliograms obtained at the Mt. Wilson Observatory between 1905 and 1984, 
and measured the area variations of plages and enhanced networks. 
The author then calibrated these area variations against the F10.7 index 
between 1947 and 1984
and constructed a proxy of the F10.7 index, extending it back to 1905.

As mentioned above, SEM EUV flux measurements are collected in multiple channels, including first-order flux and
central-order flux. 
The first-order flux measurement focuses on the ultraviolet range between 26 and 34 nm. It is derived by averaging the count rates from two channels (CH1 and CH3), which are designed to measure solar emissions in this narrow UV range.
This data is crucial for studying impacts of ionizing radiation on the Earth's atmosphere, especially the ionosphere and thermosphere. 
The data helps to understand solar variability and its influence on space weather.
The central-order flux measurement covers a broader wavelength range from 0.1 to 50 nm, capturing more of the solar spectrum's high-energy photons. It is derived from a single channel (CH2) that measures a wider range of solar emissions. This broader range of flux is used for more comprehensive solar monitoring, needed for studying the solar corona and the overall energy output of the Sun. 
The data helps to understand solar radiation's broader impacts on the Earth's environment, including climate and atmospheric chemistry.

\citet{JUDGE2000417} reported that the relative uncertainty of the SEM clone over a three-year period is only $\pm 5.5\%$ ($1\sigma$), with an absolute uncertainty of $8.5\%$ when transferring the absolute flux calibration
to the SOHO CELIAS/SEM instrument.
Our work is concerned with a long-term trend and variation in
solar EUV irradiance.
With the inspiration of the work of \citet{Foukal1998}, and the
long-term availability of SEM EUV flux measurements, we developed a Bayesian deep learning model, named SEMNet, 
for solar EUV reconstruction and
validated our model by
using CaII K images
from the Precision Solar Photometric Telescope (PSPT) to
construct SOHO/SEM EUV flux measurements
in the period between 1998 and 2014,
while minimizing the influence of solar transient events
such as solar flares and coronal mass ejections.
We then extended SEMNet through transfer learning
to reconstruct solar EUV irradiance
in the period between 1950 and 1960
using CaII K images from the
Kodaikanal Solar Observatory (KSO).
It is worth noting that
we use full-disk CaII K images rather than K line fluxes in our work because these images contain significantly more information than a single flux value. The images preserve spatial and structural variations across the solar disk, 
allowing SEMNet to learn features such as the size, brightness, and distribution of plages and active regions, some of which contribute strongly to EUV radiation.
The spatial and structural richness enables a more accurate and physically grounded mapping from chromospheric emission to the EUV flux. 
Our deep learning model is capable of capturing such a mapping, as the convolutional architectures of the model can learn hierarchical features
hidden in subtle regional patterns \citep{10.1109/CVPRW.2014.131}. 
In contrast, scalar K line fluxes lack spatial context and restrict input variability, making a machine learning model more prone to overfitting and less capable of distinguishing spatial configurations crucial for EUV emission.

Deep learning has been used for image-to-image and image-to-index transformations in solar physics
\citep{Chatzistergos2019a, Pineci2021, Son2021, Jeong2022, 2023SoPh..298...87J}. 
For example, 
\citet{Chatzistergos2019a} reconstructed
unsigned magnetic fields from CaII K images.
\citet{Pineci2021} evaluated He I 1083 nm images for extreme EUV emission. The authors applied a convolutional neural network (CNN) model to examine the nonlinear relationships in the data. They trained and validated the CNN model using historical and contemporaneous solar disk images.
\citet{doi:10.1126/sciadv.aaw6548} presented 
another CNN model to map SDO/AIA observations to spectral irradiance measurements.
In contrast to the above work, we aim to use
SEMNet and
CaII K images to
reconstruct SEM flux measurements
with uncertainty quantification.

The remainder of this paper is organized as follows. 
Section \ref{sec:observational_data} describes the
PSPT CaII K images
and the SOHO/SEM EUV flux measurements used in our study.
Section \ref{sec:method} presents the workflow and architecture of SEMNet. 
Section \ref{sec:experiment} reports the experimental results
in which we validate SEMNet by using it to
construct the SEM EUV flux measurements
in the period between 1998 and 2014.
This section also reports the results in which
we extend SEMNet through transfer learning to reconstruct
solar EUV irradiance in the period between 1950 and 1960
using the KSO CaII K images.
Section \ref{sec:conclusion} presents a discussion and concludes the paper.

\section{Observations and Data Preparation}
\label{sec:observational_data}

We utilized a combination of observational data from
different solar instruments to train our SEMNet model for reconstructing EUV irradiance from ground-based CaII K images. 
The observational data include full-disk CaII K images from the Precision Solar Photometric Telescope (PSPT)
operated at the Mauna Loa Solar Observatory (MLSO) in Hawaii \citep{1994ASPC...68...37C, 2000svc..book...75W},
along with continuous measurements of the solar broadband EUV flux in the 0.1 - 50 nm and 26 - 34 nm wavelength ranges
obtained from SOHO/SEM.

Full-disk photographs of the Sun in the resonance K line of 
singly ionized calcium, CaII K, at
3933.67Å were first obtained in the early 1890s by 
\citet{1893MmSSI..21...68H} with the spectroheliographs. Since then, regular observations with similar instruments have been made at various sites around the world.
The potential of the full-disk CaII K observations to serve as a proxy of the magnetic field makes them very valuable for studies of the evolution of solar activity \citep{2005MmSAI..76.1018O, 2009A&A...497..273L, 2016A&A...585A..40P, doi:https://doi.org/10.1002/9781119815600.ch3, Singh_2023}. 
CaII K images provide direct information on plage and network regions on the Sun.
Through the connection to solar surface magnetic fields, they have provided an excellent source to study solar magnetism for more than a century. 
The CaII K data obtained from
PSPT are characterized by their high spatial resolution of 2 arcsec/pixel scale and high photometric accuracy, with an unprecedented 0.1\% pixel-to-pixel relative photometric precision \citep{2022FrASS...942740E}.
In our study, CaII K (393.415 nm) with a full-width-half-maximum (FWHM) of 0.25 nm was used \citep{1998SoPh..177....1E, 2022FrASS...942740E}.
These data are unique in bridging historical and modern full-disk solar observations, particularly at the CaII K wavelength. 

The Solar EUV Monitor \citep[SEM;][]{1998SoPh..177..161J}, part of the Charge, Element, and Isotope Analysis System \citep[CELIAS;][]{1995SoPh..162..441H} on board the SOHO mission, is a highly stable grating spectrometer that has provided nearly continuous absolute EUV irradiance measurements since 1995. 
SEM operates in two key bandpasses: a broad range from 0.1 to 50 nm and a narrowband channel spanning 26 to 34 nm, the latter centered on the He II 30.4 nm emission line. 
The instrument utilizes a free-standing transmission grating and silicon photodiode detectors designed for long-term stability \citep{1992SPIE.1745..123O}. Initial calibrations were performed at the National Institute of Standards and Technology/Synchrotron Ultraviolet Radiation Facility (NIST/SURF) prior to launch \citep{2002ISSIR...2..135M}. To maintain calibration, an ongoing series of sounding rocket measurements have been conducted using a clone of the SEM instrument, which is calibrated before and after each flight, alongside a neon Rare Gas Ionization Cell (RGIC) absolute detector. These calibration flights have indicated some sensitivity loss in the SEM instrument, attributed to the buildup and polymerization of hydrocarbon contaminants by solar UV radiation, which has been mitigated through a time- and wavelength-dependent degradation model.
Refinements to the SEM data processing algorithm \citep{wdjSP2014}, especially during the SOHO and SDO overlapping operational period, have further improved the data accuracy, bringing the SEM irradiance measurements to a better agreement with those of the SDO/EVE instrument \citep{2012SoPh..275..115W}.

To ensure a consistent construction period, we selected the time frame
between 1998 and 2014,  
during which both CaII K images and SOHO SEM flux data (in the 0.1 - 50 nm and 26 - 34 nm wavelength ranges) are available. 
Specifically, we collected PSPT CaII K images
from the Laboratory for Atmospheric and Space Physics
(LASP)\footnote{\url{https://lasp.colorado.edu/pspt_access/}} 
at the University of Colorado Boulder, 
between March 1998 and December 2014.
We downloaded the SOHO/SEM daily average
flux measurements from
the LASP Interactive Solar IRradiance Datacenter (LISIRD).\footnote{\url{https://lasp.colorado.edu/lisird/data/soho_sem_P1D}}
On each day, there are one or more PSPT CaII K images corresponding to a pair of SOHO SEM flux readings in the two wavelength ranges.
Furthermore, since the CaII K images are appropriate for studying the
long-term (more than 27 days)
evolution of the least energetic part of the EUV spectrum
but are less effective for studying short-term variations
\citep{2008AdSpR..42..903D}, 
we excluded data samples having
anomalous SEM flux measurements
from our study,
where the anomalies may have been caused by
solar transient events.
The reason for doing so is that including those anomalies, totaling 16 with flux values significantly higher than their neighboring flux values, 
may cause our model to pay attention to the anomalies (rare short-term variations), 
which would weaken the model's ability to capture common long-term patterns in the data \citep{Aggarwal2016}.

In training our model here, we built training samples by
selecting CaII K images from PSPT taken at different timestamps on the same day, 
along with their corresponding SEM flux measurement in an EUV band on that day,
which was used as the ground-truth label for the multiple PSPT CaII K images on the day.
During testing, we randomly selected a single
PSPT CaII K image on each day and used it as a test sample. 
The training set contains
samples from January to June in the years
between 2000 and 2013,
with a total of 12,343 training samples. 
The test set contains samples from the years 1998, 1999, and 2014, 
as well as from July to December between 2000 and 2013,
totaling 1,690 test samples. 
The validation set contains 500 samples randomly chosen from the same periods as the test set but with different timestamps.
The samples in the training, test, and validation sets span a wide range of solar activity from solar minima to solar maxima between 2000 and 2013.
Note that the training set and the test set are disjoint, 
and hence our SEMNet model can
make inferences on the data that it has never seen during training.
Original PSPT CaII K images, 
with a resolution of $2048 \times 2048$ pixels, 
were resized to $256 \times 256$ pixels to improve training time, although this reduction in size did not affect model performance.
Due to different scales in the data,
we applied normalization to each PSPT CaII K image
by dividing each pixel value in the image by 1000
and to each SEM flux measurement by multiplying the measurement by 1000.

\section{Methodology} 
\label{sec:method}

\subsection{Overview of Our Approach}

\begin{figure}
\centering\includegraphics[width=0.9\linewidth]{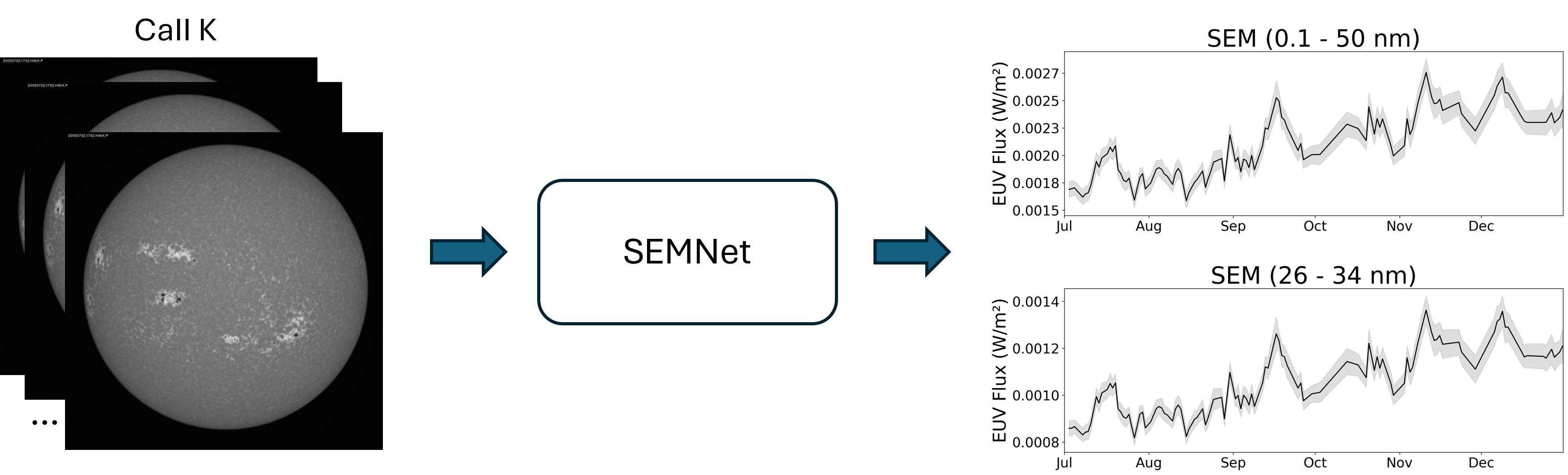}
\caption{
Illustration of the testing of SEMNet.
The trained SEMNet model takes as input individual 
256 $\times$ 256 full-disk 
CaII K images and 
produces as output 
corresponding SEM flux measurements
in the 0.1 - 50 nm and 26 - 34 nm wavelength ranges, where
the shaded areas represent estimated uncertainties.
}
\label{model_pipeline}
\end{figure}

SEMNet is designed to capture the hidden relationships between
CaII K images and SEM EUV fluxes. 
Inspired by the Monte Carlo dropout framework (MC dropout) proposed by \citet{pmlr-v48-gal16} and used in \citet{jiang2021ApJS}, we develop a ResNet-like architecture \citep{2016cvpr.confE...1H}
with MC dropout to predict SEM flux measurements with uncertainty quantification. 
Specifically, stochastic dropouts are applied, and the model output can be approximately treated as a random sample generated from the posterior predictive distribution. 
Consequently, the uncertainty of the model and the uncertainty of the data can be estimated by calculating the variance of the predictions of the model over multiple repetitions.

During training, SEMNet takes as input pairs of
CaII K images
and their corresponding SEM flux measurements in the
0.1 - 50 nm and 26 - 34 nm wavelength ranges.
The SEM flux measurements are used as ground-truth labels
to guide the learning process in which the
weights of the neurons in SEMNet are optimized.
During testing/inference, 
the trained SEMNet model takes individual $256 \times 256$ full-disk 
CaII K images as input and produces as output SEM flux measurements in the two wavelength ranges with uncertainties. 
Figure \ref{model_pipeline} explains how the trained model works during the testing.

\subsection{The SEMNet Model}

\begin{figure}
\centering\includegraphics[width=0.99\linewidth]{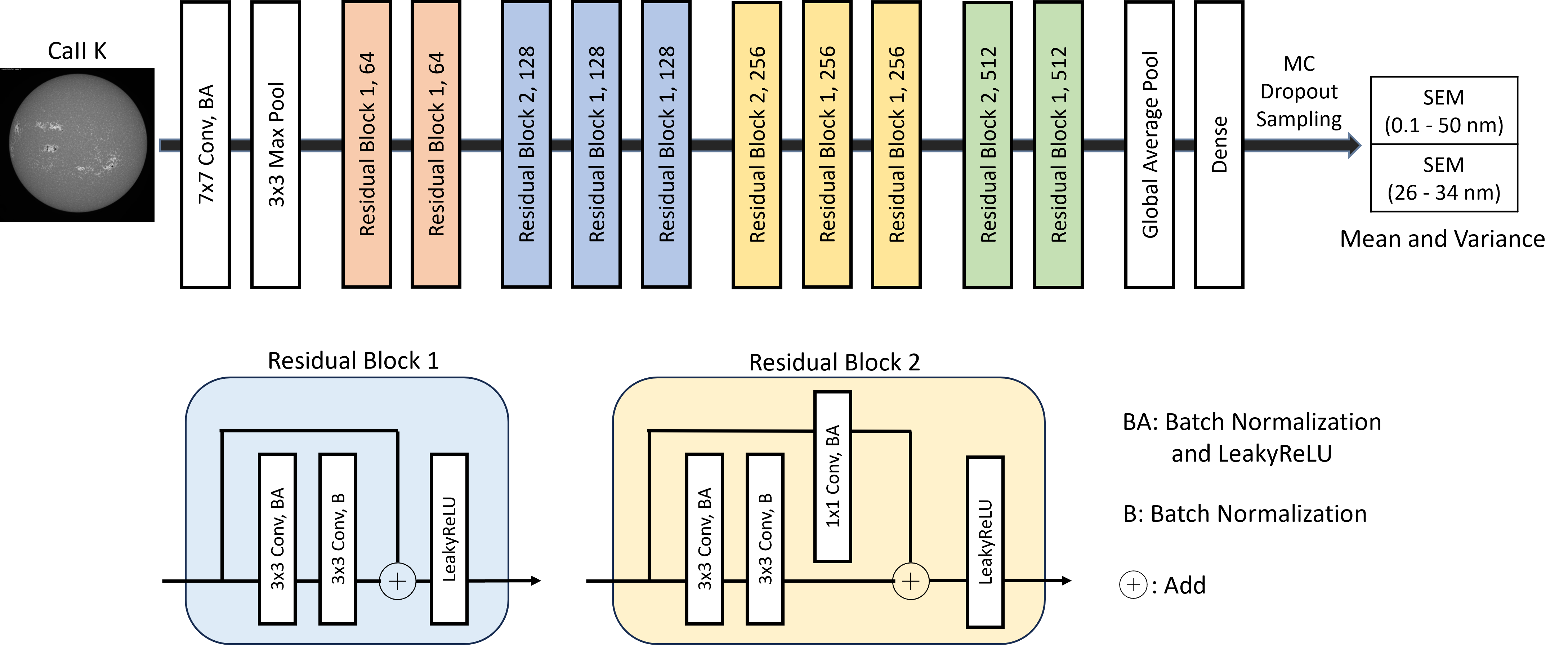}
\caption{Illustration of the architecture of SEMNet, 
which is a ResNet-like model with uncertainty quantification.}
\label{model_achitecture}
\end{figure}

Figure \ref{model_achitecture} presents the architecture of SEMNet, 
which is a ResNet-like model \citep{2016cvpr.confE...1H} 
with two types of residual blocks. 
These residual blocks are designed to learn residual functions 
with respect to the input layer, 
which mitigate the problem of vanishing gradients and accelerate convergence.
Residual block 1 consists of two convolutional layers. 
The first convolutional layer has
a kernel size of $3 \times 3$ 
with a stride of 1. 
This layer is followed by batch normalization and a LeakyReLU activation with a slope of 0.2, which maintains non-linearity while reducing the risk of dying neurons.
The second convolutional layer also employs
a kernel size of $3 \times 3$ with a stride of 1, 
ensuring dimensional consistency across the feature maps. 
Batch normalization is
applied post-convolution.
A shortcut connection is used to add
the input of this residual block to
the output of the second convolutional layer. 
The final output is activated using LeakyReLU.
Residual block 2 contains two convolutional layers
similar to those of residual block 1.  
However, unlike residual block 1, residual block 2 has a stride of 2
in the first convolutional layer to enable downsampling. 
A shortcut connection is used, which undergoes a $1 \times 1$ convolutional transformation, to add the
input of this residual block to the
output of the second convolutional layer. 
This $1 \times 1$ convolution in
residual block 2 ensures a dimension match during addition.
Like residual block 1, the final output of residual block 2 is activated using LeakyReLU.
Our SEMNet model alternates between the two types of residual blocks: residual block 1 (without downsampling) and residual block 2 (with downsampling), 
as in ResNet \citep{2016cvpr.confE...1H}.
This design allows the model to retain high spatial dimensions in the earlier layers, which is important for capturing fine-grained features in CaII K images, while progressively reducing spatial dimensions and increasing the depth of features in the later (deeper) layers to encode higher-level representations. The use of shortcut connections throughout the model helps stabilize training and mitigate vanishing gradient issues.

We feed the SEMNet model with a \(256 \times 256\) CaII K image.
The model produces as output a SEM flux measurement
in the 0.1 - 50 nm and 26 - 34 nm wavelength range, respectively,
along with uncertainty estimates.
The input image first passes through a convolutional layer with 64 filters, 
each having a kernel size of \(7 \times 7\), 
with a stride of 2. 
This initial convolution is followed by batch normalization and LeakyReLU 
with a slope of 0.2. 
A \(3 \times 3\) max-pooling layer with a stride of 2 further
downsamples the feature maps, 
capturing key spatial structures from the input. 
The network then goes through four residual block stages,
with 64, 128, 256, and 512 filters, respectively.
Following the residual stages, a global average pooling layer is used to reduce the spatial dimensions of the feature maps. This layer significantly reduces the number of parameters, enhancing generalization and providing translation invariance to the network. 
To avoid overfitting and enable the Monte Carlo (MC) dropout sampling technique \citep{jiang2021ApJS}, 
a dropout layer with a rate of 0.2 is applied, ensuring regularization by randomly deactivating neurons during training and inference. 
The final layer is a fully connected dense layer with two output units
designed for the regression task at hand. A linear activation function produces the output for the SEM flux in the 0.1 - 50 nm and 26 – 34 nm wavelength range,
respectively.

We use the Adam optimizer \citep{Goodfellow-et-al-2016} for model training, 
which minimizes the Huber loss \citep{DBLP:conf/cvpr/Barron19},
defined as:
\begin{equation}
L_{\delta}(y, \hat{y}) = 
\begin{cases} 
\frac{1}{2}(y - \hat{y})^2 & \text{if } |y - \hat{y}| \leq \delta \\
\delta \times (|y - \hat{y}| - \frac{1}{2}\delta) & 
\text{otherwise}
\end{cases}
\end{equation}
where \( y \) is the observed SEM flux value, \( \hat{y} \) is the predicted SEM flux value, and \( \delta \) is the threshold,
with a default value of 0.5,
that determines the transition between the quadratic loss and the linear loss. 
This loss function reduces the sensitivity to outliers compared to the
traditional MSE loss \citep{Goodfellow-et-al-2016}, 
which makes it well suited to handle the inherent variability of the data. 

\subsection{Uncertainty Quantification}

A Bayesian neural network combines
a neural network with a probabilistic framework for uncertainty quantification provided by Bayesian statistics. 
By placing a prior to the network parameters \( W \), 
we aim to find the posterior distribution of \( W \) 
given the training data \( D \), i.e., pairs of CaII K images and their corresponding SEM EUV flux measurements.
This process is referred to as posterior inference. 
In theory, achieving exact posterior inference in a deep network is computationally intractable \citep{pmlr-v48-gal16}, and modifying its architecture for a full Bayesian treatment can be impractical. As a result, approximate methods that preserve the original model structure are often preferred in practice.
Inspired by the approach suggested in the literature \citep{pmlr-v48-gal16, Mobiny2021}, 
we adopt the Monte Carlo (MC) dropout sampling technique to approximate the uncertainty of the model.
The algorithm works as follows: given an input CaII K image \(x\), 
the neural network output is computed with stochastic dropouts applied to the network. 
Specifically, each hidden unit is randomly dropped with a fixed probability \(p\). 
This stochastic feed-forward process is repeated \(T\) times, 
producing a set of outputs \(\{ \hat{y}_{(1)}, \ldots, \hat{y}_{(T)} \}\). 
These samples approximate the draws from the posterior predictive distribution \citep{pmlr-v48-gal16}.
The variance \(\sigma^2_{model}\), which quantifies the uncertainty of the model, is then approximated using the sampling variance:
\begin{equation}
\sigma^2_{model} = \text{Var}(f^{\mbox{SEMNet}}(x)) = \frac{1}{T} \sum_{t=1}^T \left( \hat{y}_{(t)} - \bar{\hat{y}} \right)^2,
\end{equation}
where \(\hat{y} = f^{\mbox{SEMNet}}(x)\)
is the model prediction and
\(\bar{\hat{y}} = 
\frac{1}{T} \sum_{t=1}^T \hat{y}_{(t)}\) 
is the predicted mean
\citep{pmlr-v48-gal16}. 
(In our study, $T$ is set to 50 and the dropout probability \(p\) is set to 0.2.) 
Furthermore, following the adaptive approach proposed in \citet{UQ2017Uber}, 
we let \(\hat{f}^{\mbox{SEMNet}}(\cdot) \) 
be an unbiased estimator of the network, and
\( X' = \{x_1', \dots, x_V'\} \), \( Y' = \{y_1', \dots, y_V'\} \)
be the validation data where
$x_{i}'$, $1 \leq i \leq V$, is a CaII K image and
$y_{i}'$ is its corresponding SEM flux measurement in the validation set,
and $V$ is the cardinality of the validation set.
Then, we estimate the noise level during the validation phase,
known as the inherent noise, \( \sigma^2_{noise} \), via:
\begin{equation}
\sigma^2_{noise}  = 
\frac{1}{V} \sum_{v=1}^V \left( y_v' - \hat{f}^{\mbox{SEMNet}}(x_v') \right)^2. 
\end{equation}
The final prediction uncertainty, $\sigma^2_{pred}$, is defined as $\sigma^2_{model} + \sigma^2_{noise} $. 
Once the prediction uncertainty is determined, 
we can construct an approximate \(\alpha\)-level prediction interval,
($\bar{\hat{y}} - z_{\alpha/2} \times \sigma, \; \bar{\hat{y}} + z_{\alpha/2} \times \sigma$), 
for the model prediction where
\(\sigma\) is
$\sqrt{\sigma^2_{pred}}$ and
\(z_{\alpha/2}\) is the upper \(\alpha/2\) quantile of a standard normal distribution. 
For example, when \(z_{\alpha/2} = 2\), it approximates a confidence interval of 95\%, 
which is widely used to assess
the reliability of a prediction \citep{gelman2013bayesian}. 

\section{Experiments and Results} 
\label{sec:experiment}

\subsection{Evaluation Metrics}
\label{MagNet_sec: metrics}

We adopt three metrics, namely 
the root mean square error 
\citep[RMSE;][]{gmd-7-1247-2014}, 
the mean relative error \citep[MRE;][]{doi:10.1126/sciadv.aaw6548} and the
coefficient of determination \citep[R$^{2}$;][]{10.7717/peerj-cs.623}
to quantitatively evaluate the performance of SEMNet. 
Furthermore, we use empirical coverage \citep[EC;][]{BREIDENBACH2016274}
to evaluate the uncertainty estimation performance of SEMNet.
The first metric is defined as:
\begin{equation} \label{eq:RMSE}
\text{RMSE} = \sqrt{\frac{1}{n} \sum_{i=1}^{n} (\bar{\hat{y}}_i - y_i)^2},
\end{equation}
where $n$ is the number of samples in the test set 
that contains one CaII K image/sample per day,
\(\bar{\hat{y}}_i\) is the predicted mean
and \(y_i\) is the observed value for the $i$th test sample. 
This metric calculates the square root of the average squared difference between the predicted mean value
and the observed value. 
A smaller RMSE signifies a better fit of a model to the data
with better model performance.

The second metric is defined as:
\begin{equation} \label{eq:MRE}
\text{MRE} = \frac{1}{n} \sum_{i=1}^{n} \left|\frac{\bar{\hat{y}}_i - y_i}{y_i}\right|
\times 100\%,
\end{equation}
which calculates the average relative difference between the
predicted mean \(\bar{\hat{y}}_i\)
and the observed value \(y_i\). A smaller MRE indicates better model performance.

The third metric is defined as:
\begin{equation} \label{eq:R2}
\mbox{R}^2 = 1 - \frac{\sum_{i=1}^{n}(\bar{\hat{y}}_i - y_i)^2}{\sum_{i=1}^{n}(y_i - \bar{y})^2},
\end{equation}
which measures the strength of the relationship between the 
predicted mean \(\bar{\hat{y}}_i\)
and the observed value \(y_i\) 
where \(\bar{y} = 
\frac{1}{n} \sum_{i=1}^n {y}_i\).
The metric value ranges from \(-\infty\) to 1, with a higher value indicating better model performance.

The fourth metric refers to the proportion of observed values that fall within a specified prediction interval. 
It is used to evaluate the reliability of a model's uncertainty estimates, which is defined as:
\begin{equation} \label{eq:EC}
\text{EC} = \frac{1}{n} \sum_{i=1}^n \mathbb{I} \left( y_i \in \left[ \bar{\hat{y}}_i -  z_{\alpha/2} \times \sigma_i, \bar{\hat{y}}_i + z_{\alpha/2} \times \sigma_i \right] \right)
\times 100\%,
\end{equation}
where \( \sigma_i \) is the predicted standard deviation for the $i$th test sample 
and \( \mathbb{I}(\cdot) \) is the indicator function, which is equal to 1 if the inside condition is true and 0 otherwise. For example, a confidence level of 95\%, with $z_{\alpha/2}$ = 2, 
corresponds to a prediction interval of
\([ \bar{\hat{y}}_i - 2 \sigma_i, 
\bar{\hat{y}}_i +2 \sigma_i ]\). 
Intuitively, EC represents the coverage rate. For instance, a coverage rate of 99\% means that 99\% of observed/true values fall within the prediction interval.

It should be pointed out that RMSE provides an overall assessment of the magnitude of prediction errors \citep{2022Univ....8...30Z}, while MRE offers a normalized measure of prediction errors relative to true EUV fluxes \citep{doi:10.1126/sciadv.aaw6548}. 
The two metrics are highly correlated and complementary to each other: RMSE reflects the absolute difference, whereas MRE shows the relative difference between predicted and
true EUV fluxes.
R$^{2}$ indicates how well a model explains the variance in true EUV fluxes \citep{rs17101720}. 
EC quantifies how well the specified prediction interval
covers true EUV fluxes \citep{BREIDENBACH2016274}. 
Together, these four metrics help us to better understand how accurate
a model is from an absolute and relative point of view, how well the model fits the data, and how reliable the model's uncertainty estimates are.

\subsection{Performance Assessment and Comparison}

We conducted a series of experiments to evaluate the performance of SEMNet and
compare it with related methods, including
ANet3 \citep{doi:10.1126/sciadv.aaw6548}, 
EfficientNetB0 \citep{pmlr-v97-tan19a}, and
the Vision Transformer 
\citep[ViT;][]{DBLP:conf/iclr/DosovitskiyB0WZ21}.
ANet3 was used in previous work to map AIA images to EUV fluxes \citep{doi:10.1126/sciadv.aaw6548}.
EfficientNetB0 and ViT are well-known models for image classification.
To adapt the related models to our work, 
we replaced the final classification layer of each model with a dense layer comprising two neurons, activated by a linear activation function.
Unlike SEMNet, which produces
a predicted mean and predicted variance,
the three related methods predict a single EUV flux measurement
for each wavelength range given an input PSPT CaII K image.
We chose the Adam optimizer for all three methods
to achieve the best performance and the fastest convergence.
All methods used the same training, validation and test sets as described in Section
\ref{sec:observational_data}.

\begin{table}[t]
	\caption{Performance Metric Values of SEMNet and Related Methods}
	\label{table: metrics}
	\centering
	\begin{tabular*}{0.9\textwidth}{@{\extracolsep{\fill}} ccccccc}
		\toprule
		\toprule
        \hspace{0.85cm}Wavelength & Metric & SEMNet & ANet3 & EfficientNetB0 & ViT \\ \midrule
		\multirow{4}{*}{\begin{tabular}[c]{@{}c@{}}0.1 – 50 nm\end{tabular}}
		& RMSE & \textbf{0.000200} & 0.000268 & 0.000534 & 0.000890  \\
		& MRE & \textbf{6.66$\%$} & 8.23$\%$ & 17.5$\%$ & 28.7$\%$  \\  
		& R$^{2}$ & \textbf{0.926} & 0.867 & 0.471 & $-0.466$  \\
		& EC($\mu$ $\pm$ 2$\sigma$) & \textbf{99.53$\%$} & --- & --- & ---  \\
		\midrule
		\multirow{4}{*}{\begin{tabular}[c]{@{}c@{}}26 – 34 nm \end{tabular}} 
		& RMSE & \textbf{0.000097} & 0.000129 & 0.000219 & 0.000400  \\
		& MRE & \textbf{6.17$\%$} & 7.94$\%$ & 14.7$\%$ & 26.04$\%$  \\  
		& R$^{2}$ & \textbf{0.924} & 0.866 & 0.612 & $-0.283$  \\
		& EC($\mu$ $\pm$ 2$\sigma$) & \textbf{99.82$\%$} & --- & --- & ---  \\
		\bottomrule 
	\end{tabular*}
\end{table}

Table \ref{table: metrics} presents the experimental results
based on the samples in the test set,
where the best values are highlighted in bold.
The table includes three performance metrics,
RMSE, MRE, R\(^2\), 
and an uncertainty metric, 
EC($\mu$ \(\pm\) \(2\sigma\)), 
in two wavelength ranges, 0.1 - 50 nm and 26 - 34 nm, 
where $\mu$ denotes the predicted mean and $\sigma$ denotes the predicted standard deviation.  
It can be seen from Table \ref{table: metrics} that
SEMNet outperforms the other methods in terms of both RMSE and MRE, 
achieving the lowest MRE of 6.66\% for the 0.1 - 50 nm wavelength range and 6.17\% for the 26 - 34 nm wavelength range. 
SEMNet also achieves the highest R\(^2\) values,
with 0.926 for the 0.1 - 50 nm wavelength range and
0.924 for the 26 - 34 nm wavelength range, 
showing a strong correlation between the predicted and observed
EUV flux measurements.
ViT has the worst performance, with the highest MRE values 
of 28.7\% for the 0.1 - 50 nm wavelength range 
and 26.04\% for the 26 - 34 nm wavelength range 
and negative R\(^2\) values. 
ViT is designed to capture complex spatial relationships in an image,
which are lacking in this application.
Another reason is the limited variability of the images,
which leads to poor ViT performance \citep{doi:10.1126/sciadv.aaw6548, 9878673}.
We also performed additional experiments
using other deep learning models,
e.g. ResNet50 \citep{2016cvpr.confE...1H},
which produced worse results than SEMNet.

The uncertainty metric
EC($\mu$ ± $2\sigma$) provides valuable information on the reliability of the predictions made by SEMNet. 
This metric assesses how well prediction intervals capture true/observed values, offering a practical assessment of
SEMNet's uncertainty estimates. 
Table \ref{table: metrics} shows that
EC($\mu$ ± $2\sigma$) is 99.53\% for the 0.1 - 50 nm wavelength range
and 99.82\% for the 26 - 34 nm wavelength range, 
indicating that almost all true/observed values fall within
the prediction interval of
[$\mu$ $-$ $2\sigma$, $\mu$ + $2\sigma$].
SEMNet is conservative in the sense that the confidence level of the model is 95\%,
i.e. the model expects that 95\% of the
true/observed values fall within the prediction interval,
but the coverage rate, EC, is actually higher than 95\%.
Conservative uncertainty estimates are suitable for long-term EUV reconstruction, where identifying a trend is key.
In summary, SEMNet effectively balances accuracy (in terms of RMSE, MRE, and R$^2$) and reliability (in terms of the EC metric), making it a feasible tool
for applications that require high reconstruction accuracy and quantification of uncertainty.

\begin{figure} [!t]
\centering\includegraphics[width=0.8\linewidth]{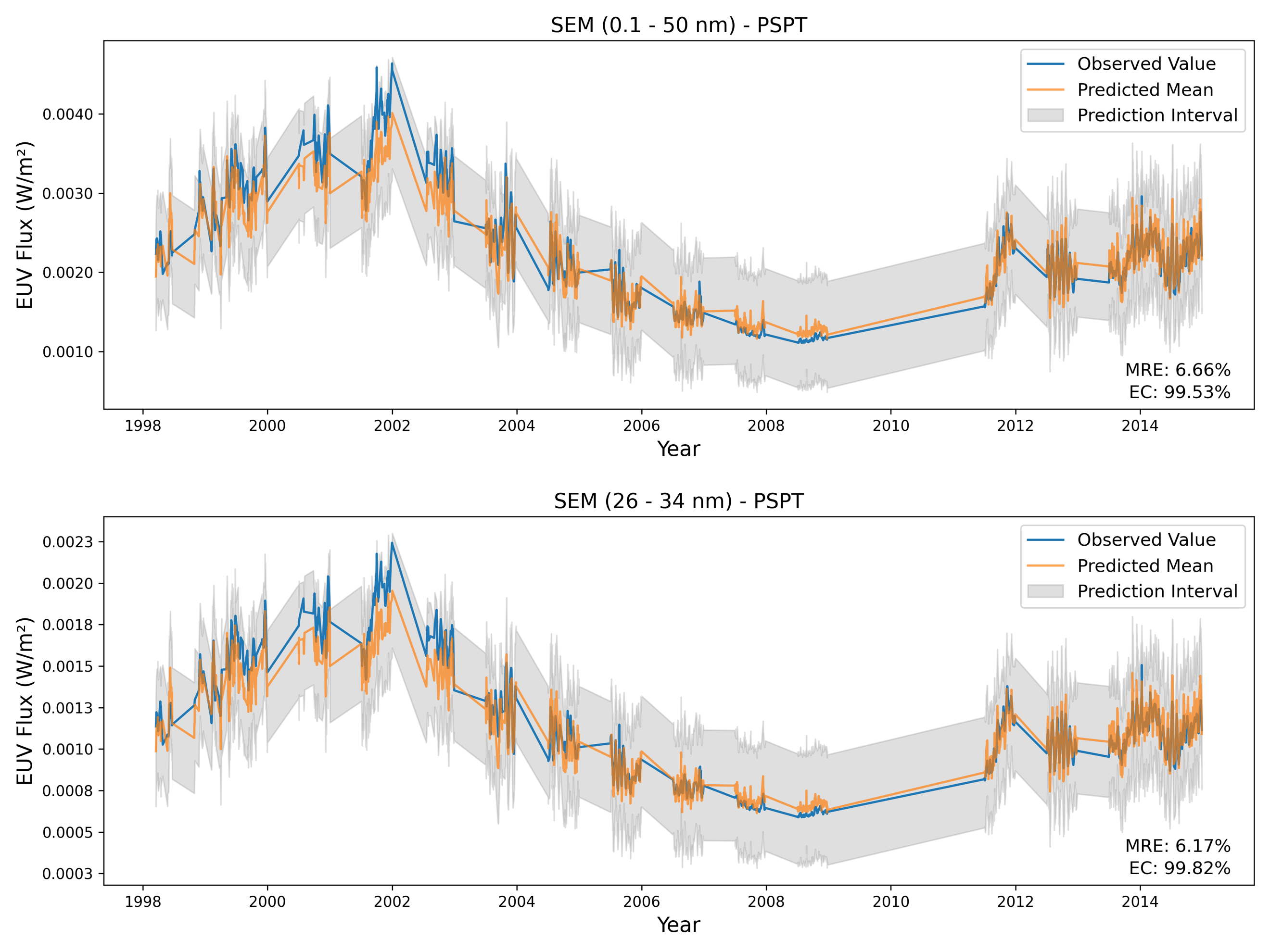}
\caption{
Observed and SEMNet-predicted EUV flux measurements 
in the period between 1998 and 2014 
based on the PSPT CaII K images in the test set.
A shaded region represents the prediction interval of
[$\mu$ $-$ $2\sigma$, $\mu$ + $2\sigma$].
Top panel: results for the 0.1 - 50 nm wavelength range.
Bottom panel: results for the 26 - 34 nm wavelength range.
}
\label{SEM_direct_comparison_all}
\end{figure}

Figure \ref{SEM_direct_comparison_all} shows the long-term evolution of observed and predicted EUV flux measurements in the period between 1998 and 2014 based on the samples in the test set. 
In the top panel of Figure \ref{SEM_direct_comparison_all}, the predicted SEM flux for the 0.1 - 50 nm wavelength range closely matches the observed SEM flux, achieving an MRE of 6.66\%. 
The shaded region represents the prediction interval of
[$\mu$ $-$ $2\sigma$, $\mu$ + $2\sigma$].
Approximately 99.53\% of the true/observed values fall within this interval, with a confidence level of 95\%.
The bottom panel of Figure \ref{SEM_direct_comparison_all} shows the predicted SEM flux
for the 26 - 34 nm wavelength range. The predicted flux aligns well with the observed data, with an MRE of 6.17\%.
Approximately 99.82\% of the true/observed values
fall within the prediction interval,
with a confidence level of 95\%.

\begin{figure} [!t]
\centering\includegraphics[width=0.99\linewidth]
{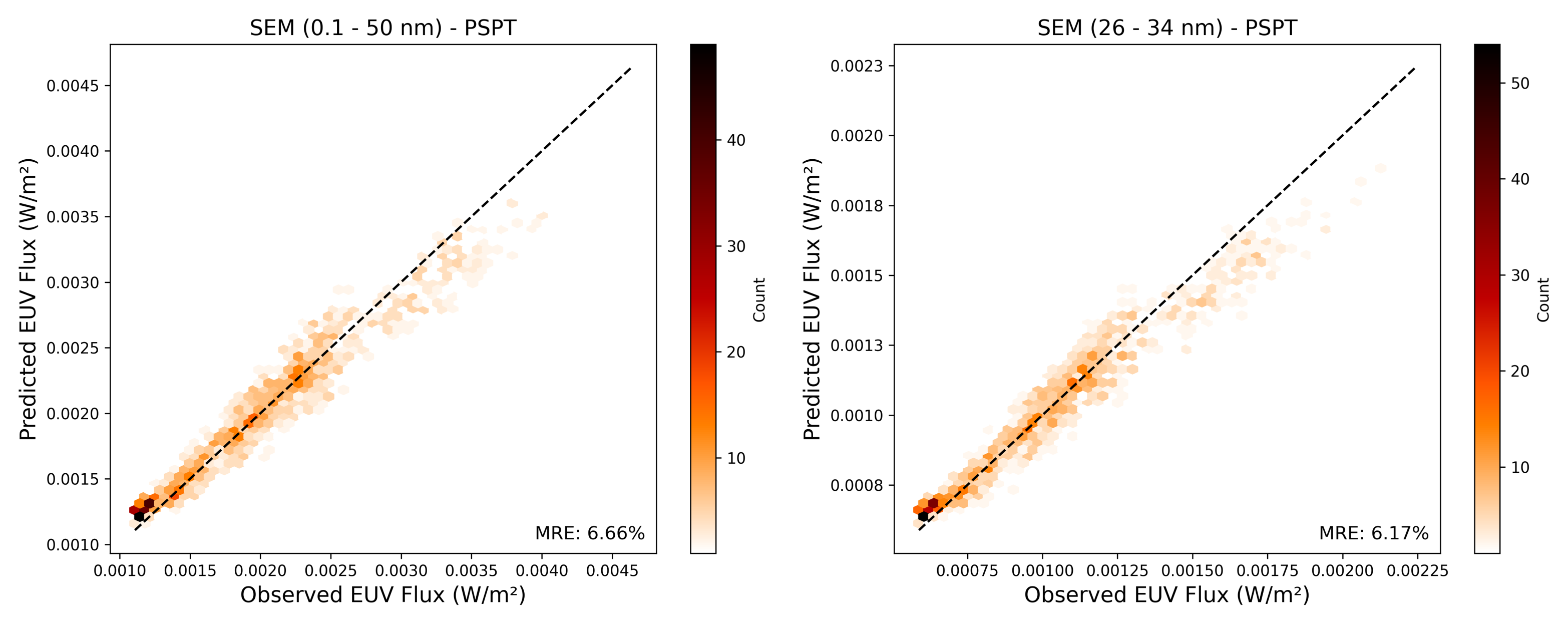}
\caption{2D histograms of observed and SEMNet-predicted 
EUV flux measurements based on the PSPT CaII K images in the test set.
Left panel: histogram for the 0.1 - 50 nm wavelength range.
Right panel: histogram for the 26 - 34 nm wavelength range.
}
\label{SEM_hist_all}
\end{figure}

Figure \ref{SEM_hist_all}
presents two-dimensional (2D) histograms of the observed and predicted EUV flux measurements
based on the samples in the test set where the 
$x$-axis ($y$-axis) in a histogram represents the observed measurements
(predicted measurements).
In the diagrams, the closer the points are scattered around the diagonal line, the better the predicted measurements are compared to the observed measurements.
It can be seen from Figure \ref{SEM_hist_all} that when the EUV flux is small, the two measurements tend to agree closely with each other.
However, when the EUV flux is large, 
the SEMNet predictions become less accurate. 
This happens probably because there are relatively few training samples in periods of strong solar activity. 
Moreover, Figure \ref{SEM_hist_all} shows that our model performs better in the 26 – 34 nm wavelength range
with an MRE of 6.17\%
than in the broader 0.1 – 50 nm wavelength range
with an MRE of 6.66\%.
This happens probably because the 26 – 34 nm wavelength range, dominated by the He II 30.4 nm line, is produced in the high chromosphere or transition region, which are layers more directly linked to CaII K emission. 
In contrast, the 0.1 – 50 nm wavelength range includes a substantial coronal contribution, which is less directly linked to CaII K emission.

\begin{figure}
\centering\includegraphics[width=0.9\linewidth]{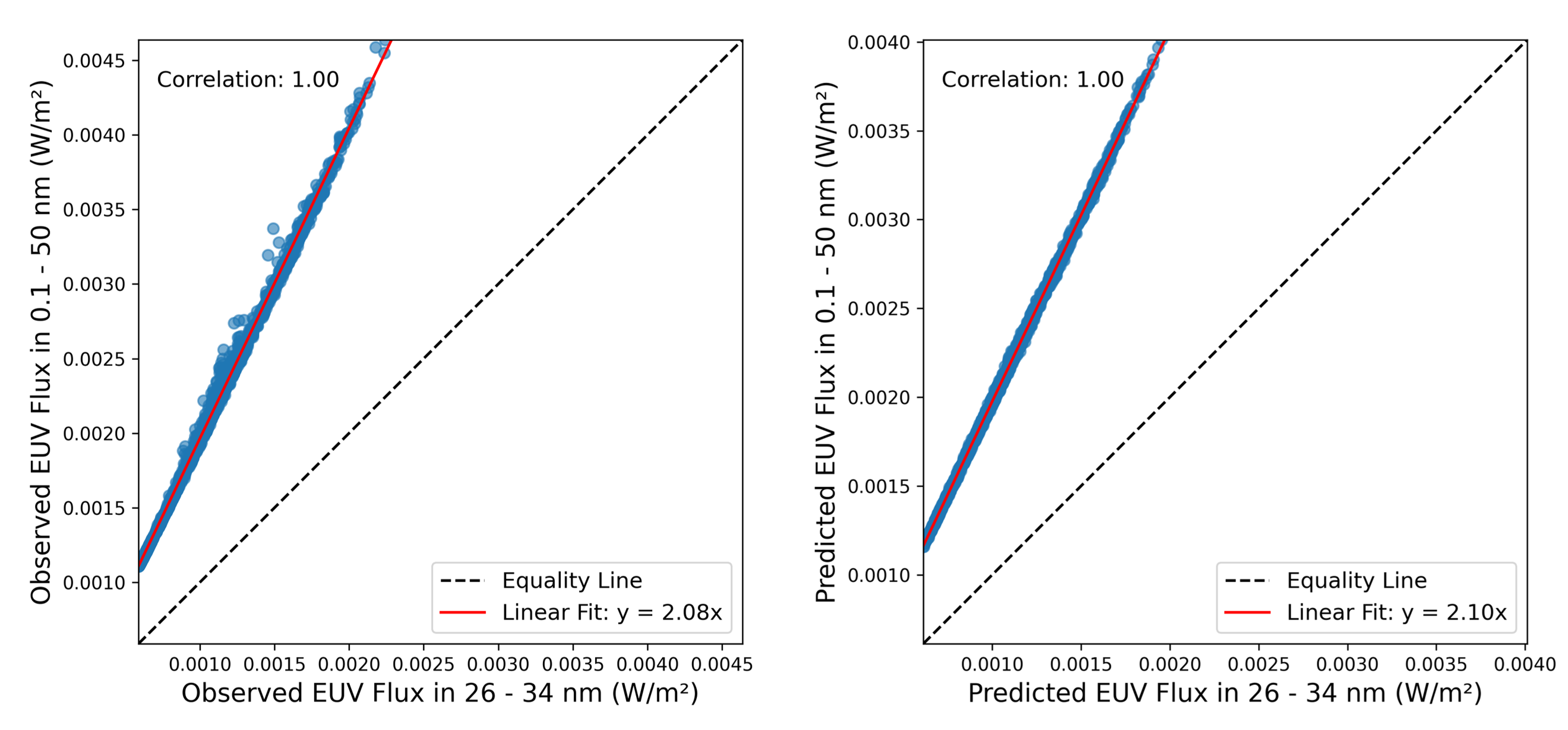}
\caption{
Scatter plots showing that the two EUV channels have similar trends in observed and SEMNet-predicted EUV flux measurements based on the PSPT CaII K images in the test set.
Left panel: scatter plot for the observed EUV flux measurements showing that there is a linear relationship, $y = 2.08x$, between the two EUV channels.
Right panel: scatter plot for the predicted EUV flux measurements showing that there is also a linear relationship, $y = 2.10x$, between the two EUV channels.
The similar trends in the data
present another level of validation of SEMNet.
}
\label{SEM_yearly_scatterplot}
\end{figure}

Figure \ref{SEM_yearly_scatterplot} 
presents scatter plots between
observed EUV flux measurements and predicted EUV flux measurements,
respectively, based on the samples in the test set.
The figure shows that
the two EUV channels have similar trends.
Specifically, there is a linear relationship, $y = 2.08x$, between the two EUV channels in the observed EUV flux measurements, 
in part because there is an overlapped range between the two EUV channels.
Furthermore, there is also a linear relationship, $y = 2.10x$, between the two EUV channels in the predicted EUV flux measurements.
Similar trends in the data
further validate the reliability of the
predictions made by SEMNet,
demonstrating the feasibility of CaII K images as a robust proxy for the long-term EUV flux, a finding consistent with that in the literature
\citep{2008AdSpR..42..903D}.

\subsection{Case Studies on the Solar Maxima in Solar Cycles 23 and 24}

\begin{figure}
\centering\includegraphics[width=0.8\linewidth]
{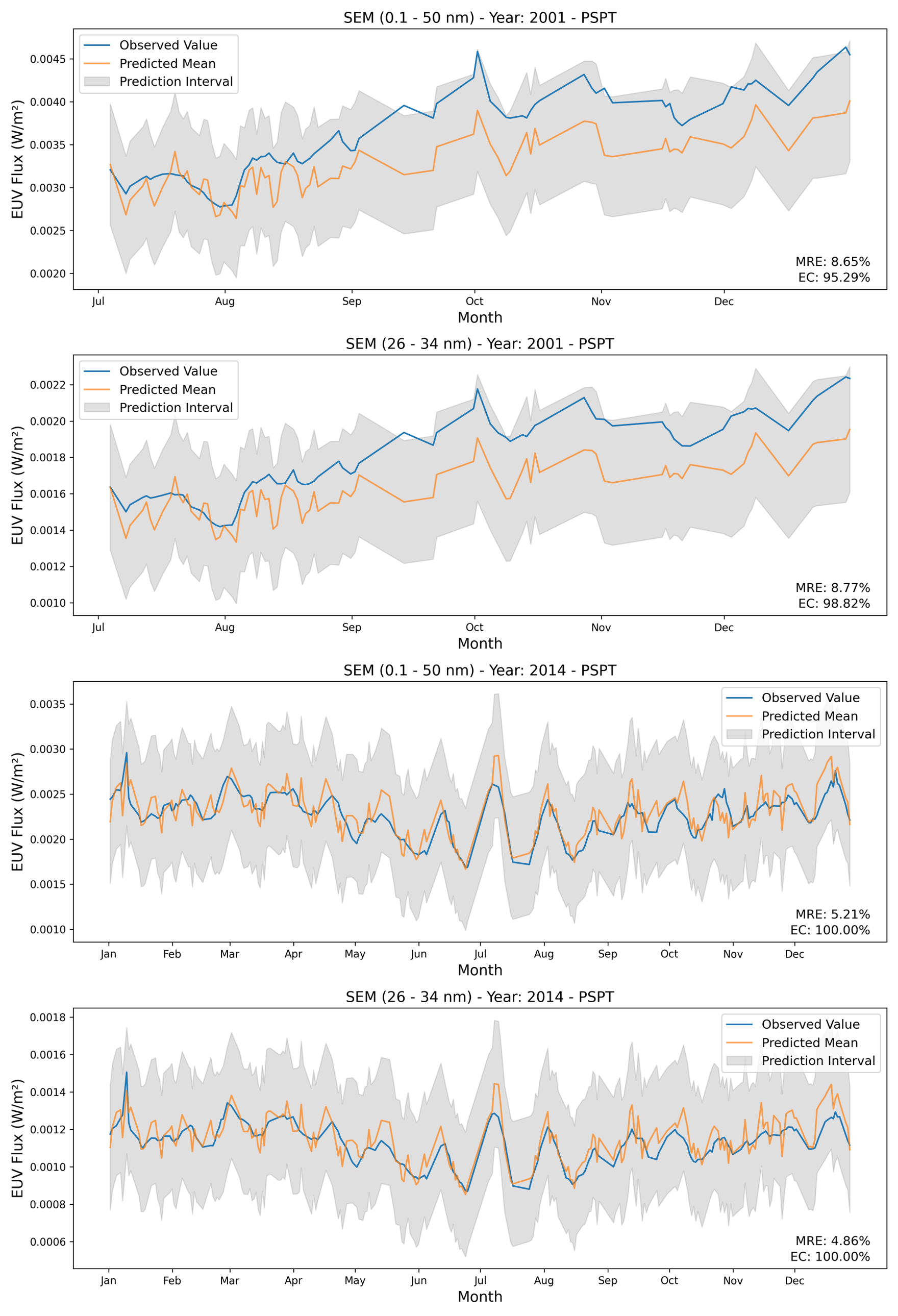}
\caption{
Observed and SEMNet-predicted EUV flux measurements during
the solar maxima in Solar Cycles 23 and 24
based on the PSPT CaII K images in the test set.
A shaded region represents the prediction interval of
[$\mu$ $-$ $2\sigma$, $\mu$ + $2\sigma$].
Top two panels: results for the two wavelength ranges 
under consideration in 2001.
Bottom two panels: results for the two wavelength ranges in 2014.
}
\label{SEM_yearly}
\end{figure}

To demonstrate how solar activity affects the predictions made by SEMNet, 
we considered the solar maximum (2001) in Solar Cycle 23 and
the solar maximum (2014) in Solar Cycle 24.
The Sun was very active in 2001 and relatively quieter in 2014.
Figure \ref{SEM_yearly} presents the results during the solar maxima.
It can be seen from the figure that our SEMNet model performs reasonably well in these periods.
Specifically, in 2001, the model achieved an MRE of 8.65\% for the 0.1 – 50 nm wavelength range and 8.77\% 
for the 26 – 34 nm wavelength range, as shown in the top two panels of Figure \ref{SEM_yearly}. 
In 2014, the model achieved an MRE of 5.21\% for the 0.1 – 50 nm 
wavelength range and 4.86\% for the 26 – 34 nm wavelength range, as shown in the bottom two panels of Figure \ref{SEM_yearly}. 
In addition, the coverage rate, EC, provides information on the reliability of the uncertainty estimates of the model. 
During 2001, which is a highly active period, 
EC($\mu$ ± $2\sigma$) reached 
95.29\% for the 0.1 – 50 nm wavelength range and
98.82\% for the 26 – 34 nm wavelength range,  
showing that the uncertainty intervals capture most of the variability
in the data. During the relatively quieter period of 2014, 
EC($\mu$ ± $2\sigma$) reached 100\% coverage.

The larger deviations from observed values during the very active period (2001) of the Sun are probably due to fewer training samples for intense solar activity, 
as there may exist large variations in solar activity when the Sun is very active, while solar activity is more uniform in quieter periods. 
These results underscore the importance of uncertainty quantification, which produces interval predictions,
as opposed to point predictions, across varying solar conditions. 

\subsection{Application of SEMNet to Kodaikanal Digitized CaII K Images via Transfer Learning}
\label{sec:KSO data and experiements}

So far, we have used PSPT CaII K images to predict SEM EUV flux measurements.
Here, we use SEMNet for the digitized CaII K data obtained from the Kodaikanal Solar Observatory (KSO)\footnote{\url{https://kso.iiap.res.in/data}} in India
to show the applicability of our method to different types of CaII K images.
Historical CaII K observations at Kodaikanal were performed using a spectroheliograph with a spectral width of 0.5 nm \citep{2014SoPh..289..137P}.
KSO uses photographic plates recorded through a telescope having a 30 cm objective with an f-ratio of \(f/21\). 
The effective spatial resolution was about 2 arcsec/pixel for the majority of the documentation time. Recently, 16-bit digitization has been performed on these plates using a CCD sensor (with a pixel size of 15 microns cooled at \(-100^\circ\)C) 
to generate \(4096 \times 4096\) raw images
\citep{Chatzistergos2018,Chatzistergos2019b,Chatzistergos2021}.
We collected and preprocessed all available digitized CaII K images from KSO, 
in the period from 1996 to 2007,
where the images were corrected for center-to-limb variations.
Specifically, to reduce cross-instrument domain discrepancies, we applied preprocessing steps such as intensity normalization and histogram matching to bring the
KSO images to the same dynamic range and style as PSPT.
These images were matched with the corresponding SEM flux readings in 
the 0.1 – 50 nm and 26 – 34 nm wavelength ranges,
ensuring temporal alignment between solar images and spectral flux measurements.

We divided the total of
370 KSO CaII K images, preprocessed to $256 \times 256$ pixels, together with their corresponding SEM flux measurements, into training, validation, and test sets.
The training set contains 241 paired samples from 2000 to 2007. 
The validation set contains 50 randomly selected paired samples from 1996 to 2007.
The test set contains 79 paird samples that cover years from 1996 to 1999. 
The training, validation, and test sets are disjoint.
We adopt a transfer learning approach, where
the SEMNet model, initially trained using the PSPT data,
is fine-tuned by being retrained by the 241 paired samples of KSO data
in the training set here.
Transfer learning provides an effective way to build a new
model to specific needs without requiring substantial training data.

\begin{figure}
\centering\includegraphics[width=0.8\linewidth]{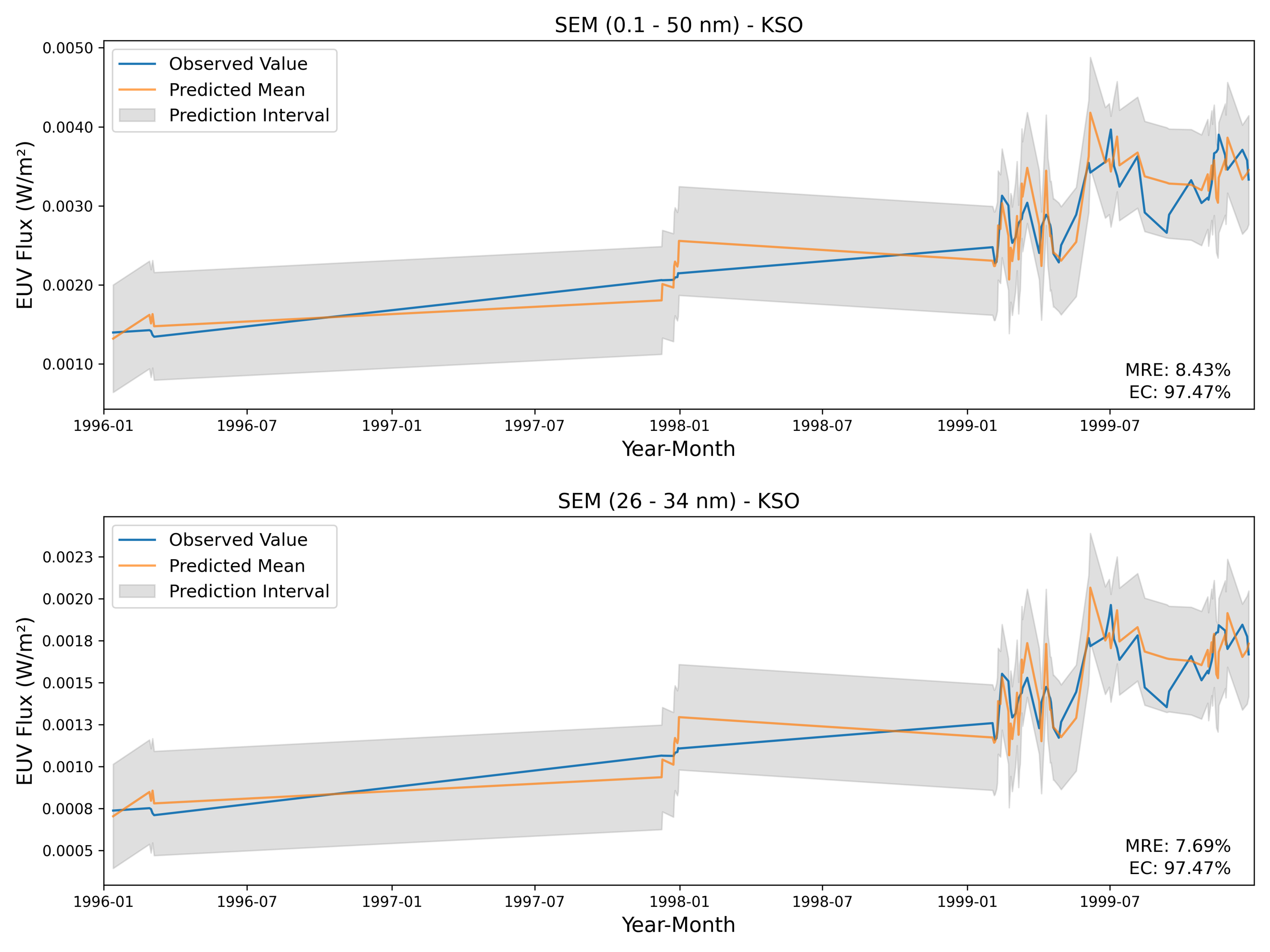}
\caption{
Observed and SEMNet-predicted EUV flux measurements 
in the period between 1996 and 1999 
based on the KSO CaII K images in the test set.
Top panel: results for the 0.1 - 50 nm 
wavelength range.
Bottom panel: results for the 26 - 34 nm wavelength range.
}
\label{SEM_KSO_1996_2000}
\end{figure}

Figure \ref{SEM_KSO_1996_2000} shows the observed and predicted EUV flux measurements
in the period between 1996 and 1999 based on the KSO CaII K images in the test set.
The fine-tuned SEMNet model works reasonably well for the KSO data. 
Specifically, the model achieves
an MRE and EC($\mu$ ± $2\sigma$) 
of 8.43\% and 97.47\% 
(7.69\% and 97.47\%, respectively)
for the wavelength range of 0.1 - 50 nm
(26 - 34 nm, respectively).
Compared to Figure \ref{SEM_direct_comparison_all},
the results based on the PSPT data with MRE of
approximately 6\%
are slightly better than those based on the KSO data
with MRE of approximately 8\%.
This is understandable given that the PSPT data benefit from more advanced modern observation technologies compared to the digitized CaII K images from KSO. 

\subsection{Reconstruction of Solar EUV Irradiance Between 1950 and 1960 Using KSO CaII K Images}

\begin{figure} [!t]
\centering\includegraphics[width=0.8\linewidth]{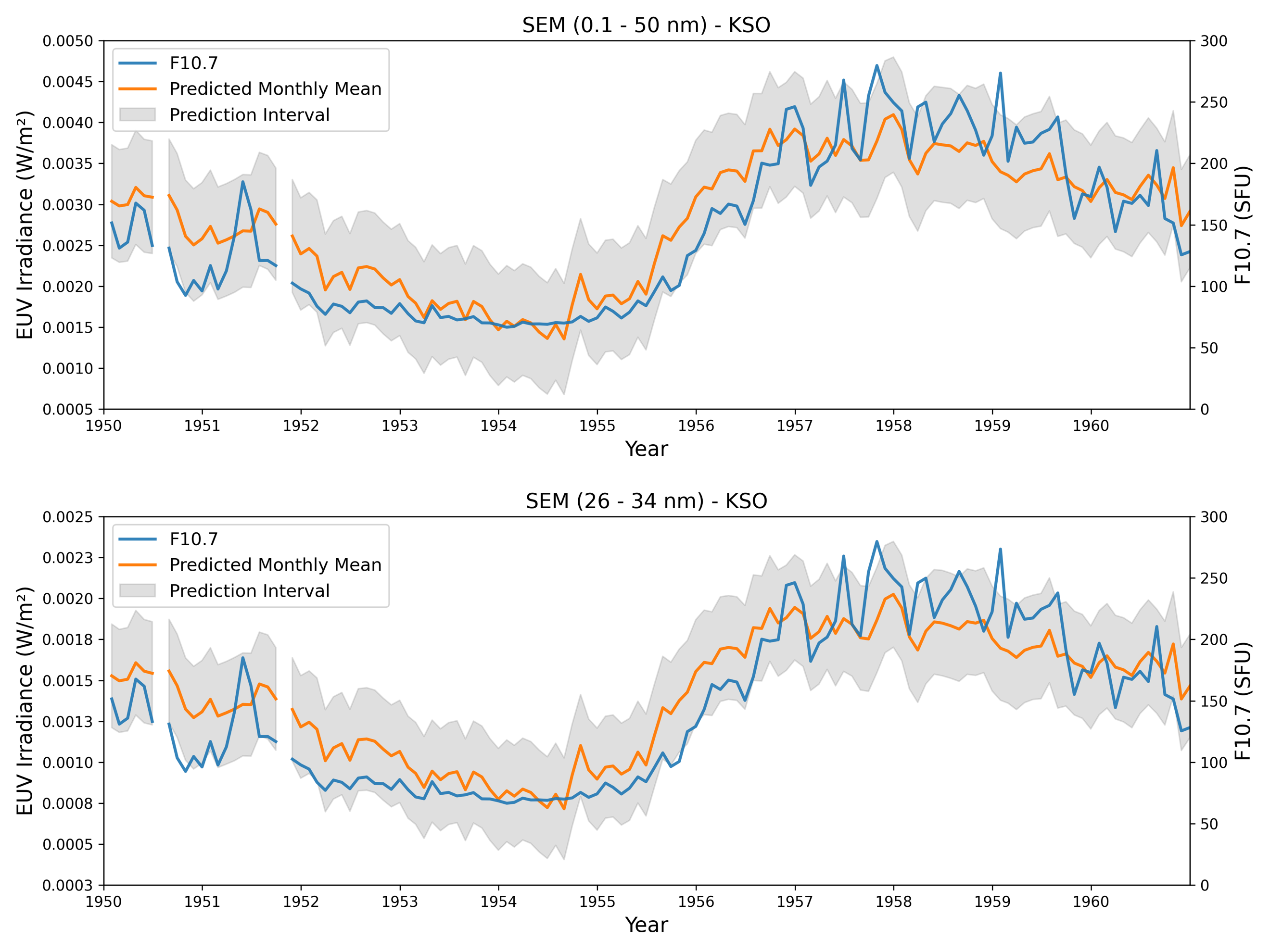}
\caption{Reconstruction of solar EUV irradiance, more 
precisely SEM EUV fluxes, in the period between 1950 and 1960 based on the KSO CaII K images
in this period.
Shown in the figure are F10.7 values, used as a reference,
and SEMNet-predicted EUV flux measurements, averaged over each month for visibility,
for the 0.1 - 50 nm wavelength range (top) 
and the 26 - 34 nm wavelength range (bottom).
Gaps on the curves are due to missing 
F10.7 values or KSO images where the two types of data cannot be temporally aligned. 
The long-term evolution trend of F10.7 is similar to that of 
reconstructed SEM EUV fluxes.}
\label{SEM_KSO_1950_1960}
\end{figure}

Here, we collect the KSO CaII K images in the period between 1950 and 1960
and attempt to use the fine-tuned SEMNet model to reconstruct
solar EUV irradiance in this period.
We considered this period because the F10.7 values,
used as a reference here, were available after 1947.
These F10.7 values were downloaded from LISIRD.\footnote{\url{https://lasp.colorado.edu/lisird/data/noaa_radio_flux}}
Figure \ref{SEM_KSO_1950_1960} presents the reconstruction results,
which are consistent with
solar activity in Solar Cycles 18 and 19. 
Specifically, the solar minimum (1954) has low EUV fluxes, while
the solar maximum (1958) has high EUV fluxes
\citep{2013JSWSC...3A..24S, 2023LRSP...20....2U}.
Because observed/true SEM EUV flux measurements are not available in this period, we present the F10.7 values as a reference in this period.
F10.7 does not reflect EUV irradiance precisely due to an additional component in gyrosynchron emission \citep{2001JGR...10610645L}, but would provide a general trend. 
Figure \ref{SEM_KSO_1950_1960} shows that the F10.7 values are in good agreement with the reconstructed
solar EUV irradiance with uncertainty estimates, especially in the solar minimum
\citep{2006JGRA..111.8304L, 2011JGRA..116.4304C}. 

\section{Discussion and Conclusions} 
\label{sec:conclusion}

The Solar and Heliospheric Observatory (SOHO) mission has provided invaluable long-term data to advance our understanding of the Sun and its influence on space weather and Earth's atmosphere. However, continuous monitoring of solar EUV irradiance remains a significant challenge.
In this study, we develop a Bayesian deep learning model (SEMNet) to predict the SEM EUV flux using CaII K images, 
focusing on the 0.1 - 50 nm and 26 - 34 nm
wavelength ranges. 
SEMNet not only provides reasonably accurate predictions but also reliably quantifies uncertainty, an essential feature in scientific settings.
Experimental results based on PSPT data
show that SEMNet consistently achieves the lowest mean relative errors (MREs), outperforming three related models (ANet3, EfficientNetB0, and ViT). 

After fine-tuning the SEMNet model via transfer learning, we then applied the fine-tuned model to predict EUV fluxes using KSO data in the period between 1996 and 1999, with reasonably good prediction results.
Moreover, we used the fine-tuned model to reconstruct solar EUV irradiance
in the period between 1950 and 1960 spanning Solar Cycles 18 and 19.
The reconstruction results are consistent with solar activity in the two solar cycles.

It should be noted that,
although our model is trained on high-quality PSPT data,
systematic errors are inevitable.
There are three causes of systematic errors.
First, the low-level fixed-pattern residuals and flat-field imperfections in the PSPT images may influence feature learning, 
particularly in low-contrast regions.
Second, the SEM EUV flux values, used as ground-truth labels in model training,
carry their own measurement uncertainties, 
especially during periods of elevated solar activity. 
These uncertainties propagate through the regression process; 
however, our prediction framework does not handle label-related uncertainties.
Third, the PSPT and KSO data differ in
many aspects, such as spatial resolution, photometric response, and dynamic range.
When applying our model, initially trained on the PSPT data, to
the KSO data through transfer learning, we applied
preprocessing steps to reduce cross-instrument domain discrepancies,
as described in Section \ref{sec:KSO data and experiements}.
Although these preprocessing steps significantly reduce discrepancies, residual inconsistencies in noise texture and stray light remain a possible source of errors.
Fine-tuning the model on the preprocessed KSO data further
improves model generalization and accuracy.
As revealed by our additional experimental study,
the errors (in terms of MRE) caused by cross-instrument domain discrepancies
are estimated to be 2.4\% for the 0.1 - 50 nm wavelength range 
and 2\% for the 26 - 34 nm wavelength range, respectively
(see Figure \ref{predictions_across_data_source} in the Appendix). 

The PSPT CaII K images have a full-width-half-maximum (FWHM) bandpass of 0.25 nm, which includes both chromospheric emission and photospheric absorption. 
To assess the impact of FWHM on the scaling relation or mapping from
CaII K images to SEM EUV fluxes, we conducted another additional experiment.
In this experiment, we downloaded PSPT CaII K Narrow Band Core (NBC) images with a FWHM of 0.1 nm from the Laboratory for Atmospheric and Space Physics (LASP).\footnote{\url{https://lasp.colorado.edu/lisird/data/pspt_cak_nbc_image_files}}
Our SEMNet model was trained on the PSPT CaII K data (FWHM 0.25 nm) from January to June in the years between 2000 and 2013 
(excluding the period of June 9-11, 2007), 
as described in Section \ref{sec:observational_data},
and then tested on three datasets, including the 
PSPT CaII K images (FWHM 0.25 nm), 
NBC images (FWHM 0.1 nm), 
and preprocessed KSO images (FWHM 0.5 nm),
respectively, in their overlapping period of June 9–11, 2007.
It was observed that the model performs better on the NBC images (FWHM 0.1 nm) than on the preprocessed KSO images (FWHM 0.5 nm);
see Figure \ref{predictions_across_FWHM} in the Appendix. 
This is understandable given that a narrower bandpass 
better isolates the chromospheric signal, 
which scales more directly with EUV emission. Another possible reason is that
the KSO data has larger cross-instrument domain discrepancies than the NBC data. 
Moreover, it was observed that the model performs better on the PSPT CaII K images 
(FWHM 0.25 nm) than on the NBC images (FWHM 0.1 nm). 
This happens probably because
the model was trained on the 
PSPT CaII K data, which might provide a favorable representation during inference.

On the basis of these experimental results, we conclude that SEMNet is a feasible tool for studying solar activity and for reconstructing solar EUV irradiance.
CaII K images are available starting 1893 and images in this spectral range have been acquired and are still acquired at several Observatories
\citep{Chatzistergos2019b, Chatzistergos2020, Chatzistergos2021, 2024JSWSC..14....9C}.
SEMNet therefore potentially allows one to extend EUV irradiance estimates 
back into the past.
\vspace*{-0.5cm}

\begin{acknowledgments}
The authors thank members of the Institute for Space Weather Sciences
for fruitful discussions.
The deep learning methods studied here
were implemented in TensorFlow.
H.J. acknowledges support from NASA grants
80NSSC24K0548 and 80NSSC25K7708.
Q.L., J.W., and H.W. acknowledge support from NSF grants
AGS-2149748, AGS-2228996, OAC-2320147, OAC-2504860, and NASA grants 80NSSC24K0548,
80NSSC24K0843, and 80NSSC24M0174.
\end{acknowledgments}

\facilities{Solar and Heliospheric Observatory, Mauna Loa Solar Observatory, Kodaikanal Solar Observatory}

\appendix

\vspace*{-0.5cm}

\section*{Supplementary Material} 

\begin{figure} 
\centering\includegraphics[width=0.8\linewidth]{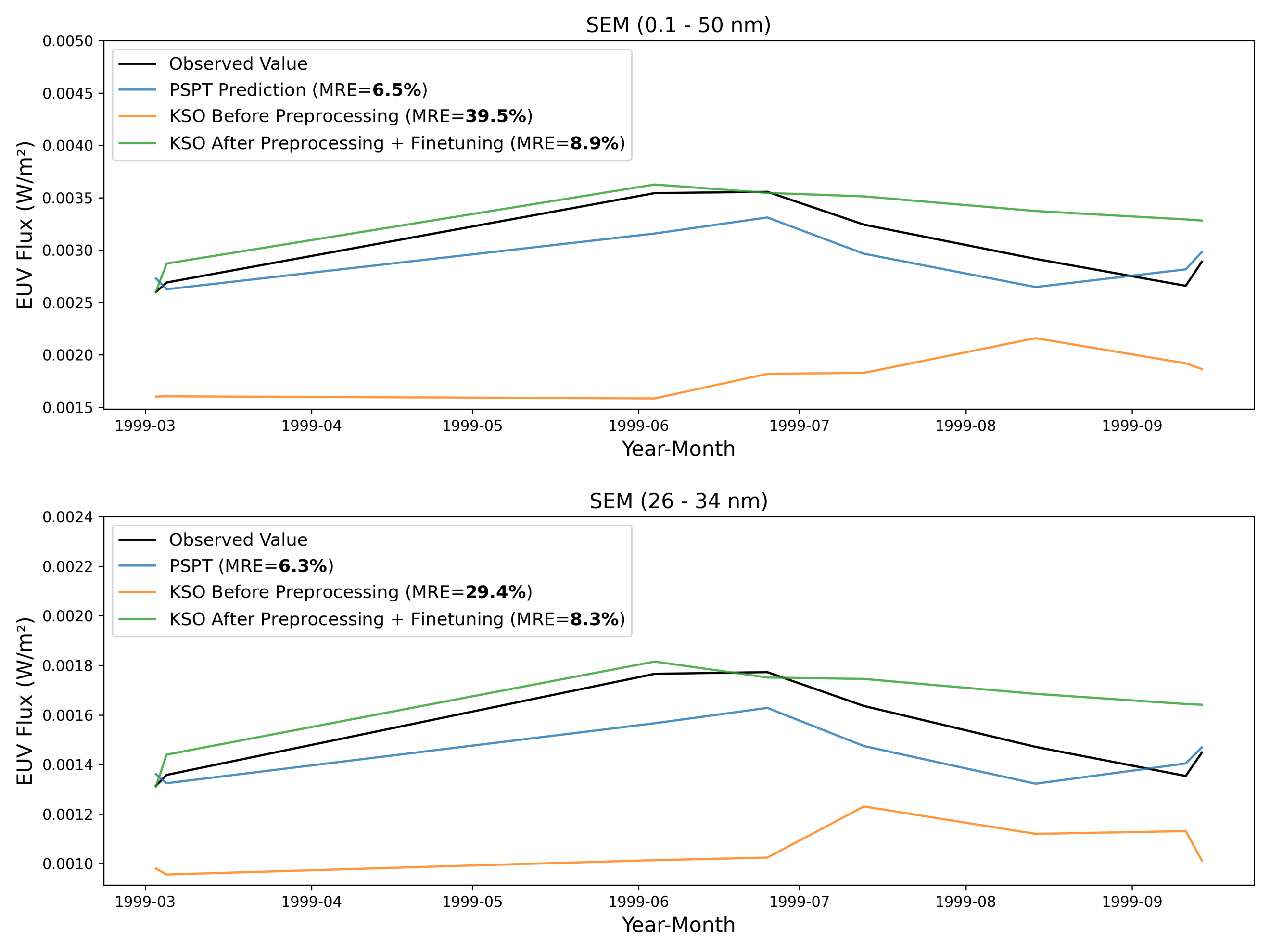}
\caption{Comparison of model predictions on
different data sources.
Shown in the figure are
observed and SEMNet-predicted EUV flux measurements on
the PSPT CaII K dataset, 
the raw KSO CaII K dataset (before preprocessing), 
and the preprocessed KSO CaII K dataset with model fine-tuning,
in their overlapping period from March to September 1999.
Top panel: results for the 0.1 – 50 nm wavelength range, with an estimated MRE of approximately 2.4\% caused by cross-instrument domain discrepancies. 
Bottom panel: results for the 26 – 34 nm wavelength range, with an estimated MRE of 2\% from cross-instrument effects.
}
\label{predictions_across_data_source} 
\end{figure}

\begin{figure} [!ht]
\centering\includegraphics[width=0.8\linewidth]{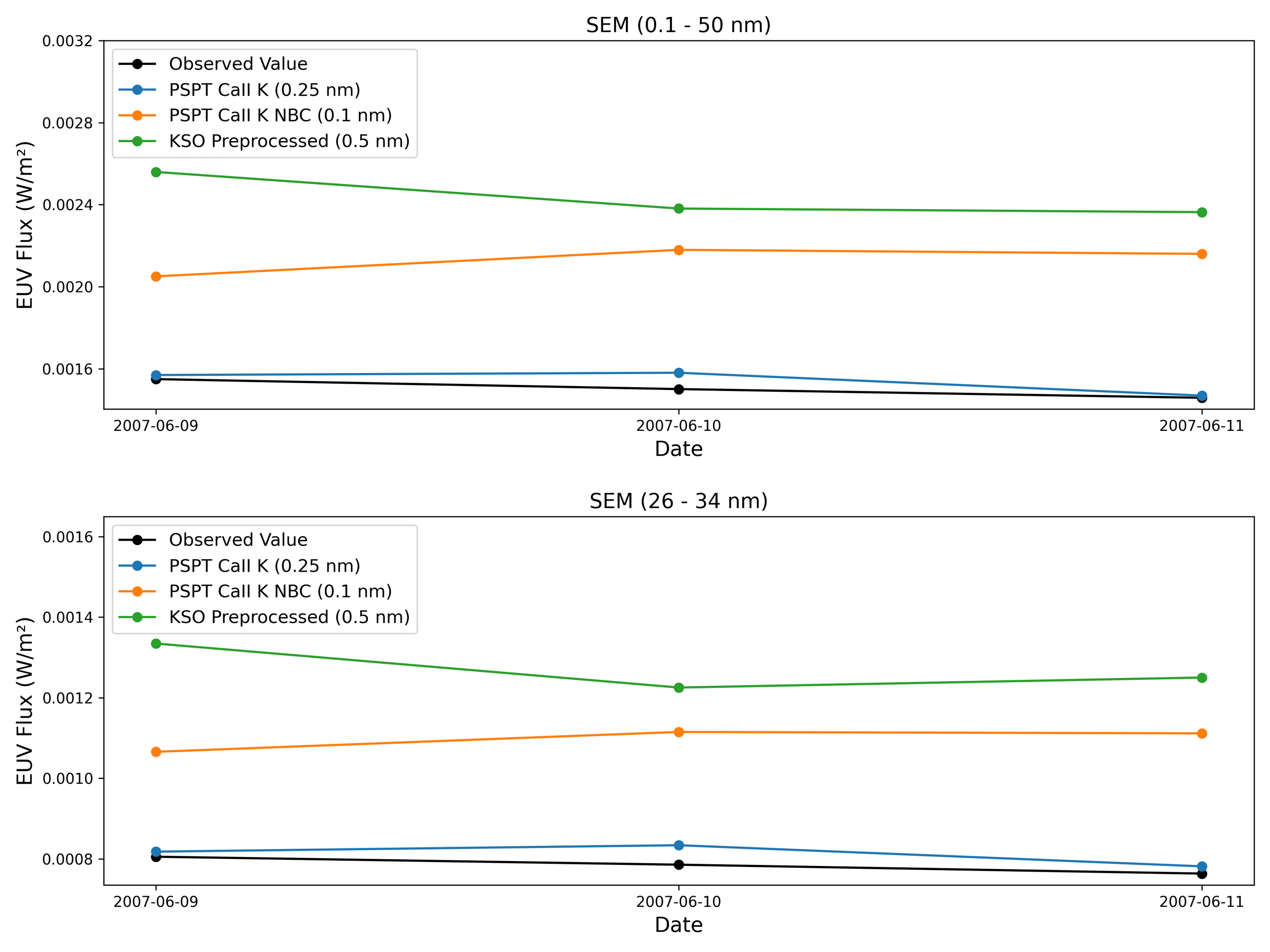}
\caption{Comparison of model predictions 
for varying FWHM.
Shown in the figure are
observed and SEMNet-predicted EUV flux measurements on
the PSPT CaII K images (FWHM 0.25 nm), 
NBC images (FWHM 0.1 nm), 
and preprocessed KSO images 
(FWHM 0.5 nm), during the period of June 9–11, 2007. 
Top panel: results for the 0.1 – 50 nm wavelength range.
Bottom panel: results for the 26 – 34 nm wavelength range.
}
\label{predictions_across_FWHM}
\end{figure}

Figure \ref{predictions_across_data_source} shows the
prediction errors, measured by the mean relative error (MRE), obtained by applying the SEMNet model, initially trained on PSPT CaII K data,
to the PSPT CaII K dataset, 
the raw KSO CaII K dataset (before preprocessing),
the preprocessed KSO CaII K dataset with model fine-tuning,
in their overlapping period from March to September 1999,
for wavelength ranges of 0.1 – 50 nm and
26 – 34 nm, respectively.
Figure \ref{predictions_across_FWHM}
compares the model predictions for varying FWHM.
The SEMNet model, trained on PSPT CaII K data, was applied separately to three datasets with different FWHM values, including the PSPT CaII K images (FWHM 0.25 nm), NBC images (FWHM 0.1 nm), and preprocessed KSO images 
(FWHM 0.5 nm), during the period of June 9–11, 2007.

\bibliographystyle{aasjournalv7}

\end{document}